# KFe$_2$Se$_2$ is the parent compound of K-doped iron selenide superconductors

Wei Li[1], Hao Ding[1], Zhi Li[2], Peng Deng[1], Kai Chang[1], Ke He[2], Shuaihua Ji[1], Lili Wang[2], Xucun Ma[2], Jiang-Ping Hu[3], Xi Chen[1,*], and Qi-Kun Xue[1,*]

[1]*State Key Laboratory of Low-Dimensional Quantum Physics, Department of Physics, Tsinghua University, Beijing 100084, China*
[2]*Institute of Physics, Chinese Academy of Sciences, Beijing 100190, China*
[3]*Department of Physics, Purdue University, West Lafayette, Indiana 47907, USA*
\* To whom correspondence should be addressed. Email: xc@mail.tsinghua.edu.cn, qkxue@mail.tsinghua.edu.cn

We elucidate the existing controversies in the newly discovered K-doped iron selenide (K$_x$Fe$_{2-y}$Se$_{2-z}$) superconductors. The stoichiometric KFe$_2$Se$_2$ with $\sqrt{2}\times\sqrt{2}$ charge ordering was identified as the parent compound of K$_x$Fe$_{2-y}$Se$_{2-z}$ superconductor using scanning tunneling microscopy and spectroscopy. The superconductivity is induced in KFe$_2$Se$_2$ by either Se vacancies or interacting with the anti-ferromagnetic K$_2$Fe$_4$Se$_5$ compound. Totally four phases were found to exist in K$_x$Fe$_{2-y}$Se$_{2-z}$: parent compound KFe$_2$Se$_2$, superconducting KFe$_2$Se$_2$ with $\sqrt{2}\times\sqrt{5}$ charge ordering, superconducting KFe$_2$Se$_{2-z}$ with Se vacancies and insulating K$_2$Fe$_4$Se$_5$ with $\sqrt{5}\times\sqrt{5}$ Fe vacancy order. The phase separation takes place at the mesoscopic scale under standard molecular beam epitaxy condition.

Similar to cuprate high temperature superconductors, superconductivity in pnictides and chalcogenides is induced by suppressing the magnetic order in their iron-based parent compounds. Therefore, understanding the parent compounds is crucial for elucidating the superconducting pairing mechanism. For most of the Fe-based superconductors, it is straightforward to identify their parent compounds. However, the newly discovered K-doped iron selenide ($K_xFe_{2-y}Se_{2-z}$, Fig. 1A) superconductor (*1-3*) has shown to be unique (*4-15*) and the parent compound of this hotly debated material remains elusive. The central issue being debated is the role of the $K_2Fe_4Se_5$ compound with $\sqrt{5}\times\sqrt{5}$ Fe vacancy order. $K_2Fe_4Se_5$ is one of the coexisting phases in $K_xFe_{2-y}Se_{2-z}$ (*10-14*) and has an antiferromagnetic order with Néel temperature up to 560 K (*5, 6*). Several experiments (*3-9*) have shown that $K_2Fe_4Se_5$ has a close connection to the superconductivity in $K_xFe_{2-y}Se_{2-z}$ and suggested that $K_2Fe_4Se_5$ is the parent compound. However, the scanning tunneling microscopy (STM) data, on the other hand, indicates that the superconducting phase in $K_xFe_{2-y}Se_{2-z}$ is the stoichiometric $KFe_2Se_2$ compound (*14*).

To clarify the controversies, we grow high-quality $K_xFe_{2-y}Se_{2-z}$ thin films by using molecular beam epitaxy (MBE). The subsequent STM measurement demonstrates that the metallic $KFe_2Se_2$ with $\sqrt{2}\times\sqrt{2}$ charge ordering is the parent compound of the superconducting phases in $K_xFe_{2-y}Se_{2-z}$. Superconductivity is induced in this parent compound by either introducing Se vacancies or spatially placing it together with the anti-ferromagnetic $K_2Fe_4Se_5$ side by side. These findings explain the apparent coexistence of magnetism and superconductivity in $K_xFe_{2-y}Se_{2-z}$: Although $K_2Fe_4Se_5$ with $\sqrt{5}\times\sqrt{5}$ Fe vacancy order is not superconducting by itself, it can give rise to superconductivity in $KFe_2Se_2$ by affecting electron doping concentration in superconducting phases.

The $K_xFe_{2-y}Se_{2-z}$ thin film was prepared on $SrTiO_3(001)$ substrate because of the small lattice mismatch. High-purity Fe (99.995%), Se (99.9999%) and K were evaporated from two standard Knudsen cells and one alkali-metal dispenser (SAES Getters), respectively. The substrate was held at 400 ºC during growth. The MBE growth of the film follows the layer-by-layer mode. To remove the extra K and Se adatoms and obtain the superconducting phase, the sample was subsequently annealed at 400 ºC for 1 hour. The STM experiments were performed in the same ultra-high vacuum system. Figure 1B shows the typical topography of a film after annealing. The step height is 0.7 nm and in good agreement with the lattice parameter of

$KFe_2Se_2$. The K atoms on the topmost layer are highly mobile and most of them can be easily desorbed during annealing; thus the film is Se-terminated. We attribute the protrusions on the surface in Fig. 1B to the residual K clusters. The STM image with atomic resolution in Fig. 1C shows a 3.9 Å × 3.9 Å square lattice consistent with the X-ray data for (001) plane of $KFe_2Se_2$ (*1*). Therefore, the surface is that of a stoichiometric $KFe_2Se_2$ single-crystalline film and no surface reconstruction is observed.

Although the lattice structure is uniform throughout the film surface, inhomogeneity in the electronic structure is clearly revealed in STM images at certain bias voltages (see Fig. 1D). Generally, the film is separated into two regions labeled by I and II in Fig. 1D. The 1×1-Se square lattice is uninterrupted when crossing the boundary of the two regions. At a bias voltage within ±100 mV, region I exhibits a $\sqrt{2}\times\sqrt{2}$ superstructure (*10, 15*) (Fig. 1E) with respect to the original Se lattice. The $\sqrt{2}\times\sqrt{2}$ charge ordering has its origin in the block antiferromagnetic state of the underlying Fe layer (*16*). In the ground state of $KFe_2Se_2$, each four Fe atoms group together to form a checkerboard pattern with antiferromagnetic order, leading to a charge density modulation on Se sites. This checkerboard phase driven by magnetic exchange coupling breaks the original symmetry of the tetragonal lattice but still retains a four-fold symmetry.

Scanning tunneling spectroscopy (STS) probes the local density of states of electrons. The STS of region I shows a 10 mV dip near the Fermi level (Fig. 1F). The same feature was also observed on the cleaved $K_xFe_{2-y}Se_2$ single crystal (*15*). The dip may stem from the $\sqrt{2}\times\sqrt{2}$ charge ordering but does not imply superconductivity because the bottom of the dip still has finite density of states and the spectrum is essentially independent of temperature from 0.4 K to 4.2 K. Therefore, region I is a nonsuperconductive metal.

We observe a different charge ordering in region II (Fig. 1G). The fret-like pattern breaks the four-fold symmetry and is visible within a bias voltage of ±60 mV. The basic building block of the pattern is a $\sqrt{2}\times\sqrt{5}$ charge density modulation (see the parallelogram in the inset of Fig. 1G). The region is divided into domains depending on the orientations of the stripes.

The STS of region II (Fig. 1H) exhibits a full energy gap centered at Fermi level and two pronounced coherence peaks, indicating that region II is superconducting with a nearly isotropic gap. The superconducting gap $\Delta$=8.8 meV is estimated by half of the energy between the two coherence peaks and in close agreement with that obtained by angle-resolved photoemission spectroscopy (ARPES) (*17-20*). A smaller gap of 7.2 meV (indicated by arrows in Fig. 1H) is also revealed in STS. Since the hole pocket at $\Gamma$ point is absent in $K_xFe_{2-y}Se_{2-z}$ superconductor, superconducting gaps must appear on the electron pockets at M point. The double-gap structure can be attributed to the two d-electron bands at M point.

Although it is not feasible to do transport measurement on the $K_xFe_{2-y}Se_{2-z}$ film at present, the occurrence of superconductivity is further supported by the response of STS to external magnetic field or magnetic defects. Similar to $KFe_2Se_2$(110) film (*14*), no magnetic vortex is observed in the superconducting state of region II. Nevertheless, the effect of magnetic field is still manifested itself by reducing the coherence peaks in STS (see Fig.S1). Stronger suppression of the coherence peaks can be achieved by magnetic defects, which locally break the time-reversal symmetry (Fig. 2). Both Fe and Se vacancies carry magnetic moment and induce bound quasiparticle states in the superconducting gap (Fig. 2B and 2D). A distinct feature of such bound states in a superconductor is that the energies of the electron-like and hole-like states are symmetric with respect to zero bias whereas their amplitudes are usually different as a result of on-site Coulomb interaction.

$KFe_2Se_2$ in region I and II has the same crystal structure but exhibits very different electronic properties. We attribute the difference to the existence of antiferromagnetic $K_2Fe_4Se_5$ insulating layer below the $KFe_2Se_2$ film in region II (see the schematic in Fig. 3A). Although STM cannot directly probe the $K_2Fe_4Se_5$ layer, which is a few nanometers below the surface, the Fe vacancy order in $K_2Fe_4Se_5$ still manifests itself by inducing a $\sqrt{5}\times\sqrt{5}$ superstructure in the local density of states of the topmost Se layer in $KFe_2Se_2$ (Fig. 3B). Such superstructure is only visible at a bias voltage of about 70 mV. The projection of interface states to the topmost surface has been observed in other systems (*21-23*) and usually happens in a sample with high quality where the mean free path of electrons is comparable to the film thickness. The existence of $K_2Fe_4Se_5$ layer below $KFe_2Se_2$ is further supported by the well-defined Morié pattern marked by arrows in the STM image in Fig. 3B. The period of the pattern is $3\sqrt{5}a_{Fe}$, where $a_{Fe}$ is the Fe-Fe distance. The Morié pattern is in excellent agreement with a simple simulation (Fig. S2), where two lattices with $\sqrt{2}\times\sqrt{5}$ and $\sqrt{5}\times\sqrt{5}$ superstructures are superimposed on each other.

STM imaging of the films indicates that the growth condition in the present work always produces a film with $KFe_2Se_2$ phase on the top and no $K_2Fe_4Se_5$ phase is exposed. The possible reason is that Se is highly volatile and the Se-rich phase in the top few layers is kinetically unstable under the growth temperature. In further investigation, fine-tuning of growth conditions may alter the surface stoichiometry and lead to a film with $K_2Fe_4Se_5$ phase on the top surface.

Phase separation between superconducting $KFe_2Se_2$ and insulating $K_2Fe_4Se_5$ along c-axis has previously been demonstrated (*10, 14*). However, not all $KFe_2Se_2$ is superconducting. Here we have shown that the non-superconductive phase of $KFe_2Se_2$ has a $\sqrt{2}\times\sqrt{2}$ charge ordering, which becomes superconducting only when it interfaces with $K_2Fe_4Se_5$ and develops a $\sqrt{2}\times\sqrt{5}$ charge ordering. For this reason, it is appropriate to identify $KFe_2Se_2$ with $\sqrt{2}\times\sqrt{2}$ charge ordering as the parent compound. Similar to the $\sqrt{2}\times\sqrt{2}$ charge ordering, the $\sqrt{2}\times\sqrt{5}$ pattern may also be the result of a specific type of magnetic ordering in the Fe layer of $KFe_2Se_2$ (one possibility is shown in Fig. S3).

There are various ways, such as strain, magnetic coupling or charge transfer, that $K_2Fe_4Se_5$ layer can regulate the electronic properties of $KFe_2Se_2$. Strain effect can be simply excluded because the lattice constants of $K_2Fe_4Se_5$ are very close to those of $KFe_2Se_2$. An analogy to cuprate high temperature superconductors suggests that the $K_2Fe_4Se_5$ layer may play the role as charge reservoir and transfer carriers into $KFe_2Se_2$ to induce superconductivity. To keep the balance of chemical valence, the $KFe_2Se_2$ phase tends to lose electrons and become hole-doped in the superconducting state. Another possibility is that the antiferromagnetic $K_2Fe_4Se_5$ may change the magnetic structure of $KFe_2Se_2$ through their exchange coupling across the interface and $KFe_2Se_2$ becomes superconducting after the original $C_4$ symmetry is broken by the magnetic interaction. In all scenarios, the interface between $KFe_2Se_2$ and $K_2Fe_4Se_5$ is a key factor. The interface in the present (001) film is smoother than that in the (110) film grown on graphene [14]. The difference may help to explain the larger superconducting gap observed in this work.

Interfacing with $K_2Fe_4Se_5$ is not the only way to induce superconductivity in the parent compound $KFe_2Se_2$. Superconductivity can also occur in a film with certain amount of Se-vacancies (with a density of about one in 10 nm$^2$). The quatrefoil-like defects (Fig. 4A) appear if the substrate temperature is raised to 430ºC during growth. The defects are attributed to Se vacancies by examining their registration with respect to the lattice. The film shows the same $\sqrt{2}\times\sqrt{2}$ superstructure as the parent compound. No sign of $\sqrt{5}\times\sqrt{5}$ Fe vacancy order has been observed. STS (Fig. 4B) at a location away from Se vacancies exhibits very similar superconducting gap to that on $KFe_2Se_2$ with $\sqrt{2}\times\sqrt{5}$ charge ordering (Fig. 1H). The coherence peaks are weaker than those in Fig. 1H because of the existence of defects. The Se vacancies carry magnetic moment, giving rise to bound states (Fig. S4) similar to those in Fig. 2D. The vacancies break the magnetic ordering in $KFe_2Se_2$ and induce superconductivity in the parent compound. The disorder-induced superconductivity exists in other iron-based superconductors as well, for example, $Ba(Fe_{1-x}Ru_x)_2As_2$ (*24*) and $BaFe_2(As_{1-x}P_x)_2$ (*25*) where Fe and As are iso-valently substituted by Ru and P respectively. Uncovering this second path leading to superconductivity indicates that it is possible to prepare a superconducting $KFe_2Se_{2-z}$ sample without $\sqrt{5}\times\sqrt{5}$ Fe vacancy order.

By demonstrating two different ways to induce superconductivity in the parent compound $KFe_2Se_2$, we have elucidated the existing controversies in K-doped iron selenide superconductors. The apparent coexistence of superconductivity and antiferromagnetism with large magnetic moment is, as a matter of fact, a "symbiotic" relationship taking place at the mesoscopic scale. These findings may open a new avenue for manipulating the superconducting properties of materials.

**References**

1. J. Guo *et al.*, *Phys. Rev. B* **82**, 180520(R) (2010).
2. A. F. Wang *et al., Phys. Rev. B* **83**, 060512(R) (2011).
3. M.-H. Fang *et al., Europhys. Lett.* **94**, 27009 (2011).
4. Z. Shermadini *et al., Phys. Rev. Lett.* **106**, 117602 (2011).
5. W. Bao *et al., Chinese Phys. Lett.* **28**, 086104 ( 2011 ).
6. V. Yu. Pomjakushin *et al., Phys. Rev. B* **83**, 144410 (2011).
7. Y. J. Yan *et al.,* http://arxiv.org/abs/1104.4941 (2011).
8. P. Zavalij *et al.*, *Phys. Rev. B* **83**, 132509 (2011).
9. W. Bao *et al.,* http://arxiv.org/abs/1102.3674 (2011).
10. Z. Wang *et al, Phys. Rev. B* **83**, 140505(R) (2011).
11. B. Shen *et al., Europhy. Lett.* **96**, 37010 (2011).
12. A. Ricci *et al.*, *Phys. Rev. B* **84**, 060511(R) (2011).
13. F. Chen *et al.*, *Phys. Rev. X* **1**, 021020 (2011).
14. W. Li *et al.*, *Nature Phys.* **8**, 126–130 (2012).
15. P. Cai *et al.*, *Phys. Rev. B* **85**, 094512 (2012).
16. W. Li, S. Dong, F. Chen, J. P. Hu, *Phys. Rev. B* **85**, 100407(R) (2011).
17. Y. Zhang *et al., Nature Mater.* **10**, 273–277 (2011).
18. T. Qian *et al., Phys. Rev. Lett.* **106**, 187001 (2011).
19. L. Zhao *et al., Phys. Rev. B* **83**, 140508(R) (2011).
20. X.-P. Wang *et al., Europhys. Lett.* **93**, 57001 (2011).
21. M. Yakes, M. C. Tringides, *J. Phys. Chem. A* **115**, 7096 (2011).
22. Y. S. Fu *et al.*, *Chinese Phys. Lett.* **27** 066804 (2010).
23. F. Chen *et al.*, http://arxiv.org/abs/1102.1056 (2011).
24. R. S. Dhaka *et al.*, *Phys. Rev. Lett.* **107**, 267002 (2011).
25. S. Jiang *et al.*, *J. Phys.: Condens. Matter* **21**, 382203 (2009).

**Figure Captions**

**Fig. 1.** STM characterization of $K_xFe_{2-y}Se_2$ films grown by MBE. (**A**) The crystal structure of $KFe_2Se_2$. (**B**) Topographic image (3.9 V, 0.02 nA, 90 nm × 90 nm) of a $K_xFe_{2-y}Se_2$ film. (**C, D**) Atomic-resolution STM topography (10 nm × 10 nm) of $KFe_2Se_2$. The two images belong to the same area, but with different bias voltages: -90 mV for (C) and 50 mV for (D). The tunneling current is 0.02 A for both. (C) shows the uniform 1×1-Se square lattice. Inhomogeneity in electronic structure is revealed in (D) with two distinct regions labeled by I and II. (**E**) The $\sqrt{2}\times\sqrt{2}$ charge ordering in region I. The imaging condition: 40 mV, 0.02 nA. (**F**) *dI/dV* spectrum (set point, 25 mV, 0.1 nA) of region I, which reveals that region I is non-superconducting. (**G**) The $\sqrt{2}\times\sqrt{5}$ charge ordering (see also inset) in region II. Imaging condition: 30 mV, 0.02 nA. (**H**) *dI/dV* spectrum (0.4 K; set point, 20 mV, 0.1 nA) showing that there is a superconducting gap opened in region II. Arrows mark the smaller gap.

**Fig. 2.** Vacancy-induced bound states in superconducting gap. (**A, B**) STM topography (40 mV, 0.02 nA) and *dI/dV* spectrum (0.4 K; set point, 25 mV, 0.1 nA) of a single Fe vacancy. (**C, D**) STM topography (-95 mV, 0.02 nA) and *dI/dV* spectrum (0.4K; set point, 25 mV, 0.1 nA) of a single Se vacancy.

**Fig. 3.** The origin of superconductivity in $KFe_2Se_2$. (**A**) Schematic showing the relationship between insulating $K_2Fe_4Se_5$ and superconducting $KFe_2Se_2$. Across the interface between $KFe_2Se_2$ and $K_2Fe_4Se_5$, the lattice structure is the same except a $\sqrt{5}\times\sqrt{5}$ Fe vacancy order in $K_2Fe_4Se_5$. (**B**) $\sqrt{5}\times\sqrt{5}$ superstructure in region II. The image belongs to the same area as Fig. 1G, but with a different bias voltage (70 mV). The dashed circles highlight some of the atoms forming the $\sqrt{5}\times\sqrt{5}$ superstructure. The Morié pattern is marked by arrows.

**Fig. 4.** Superconductivity in $KFe_2Se_{2-z}$. (**A**) STM topography (75 mV, 0.02 nA) of an area with Se vacancies. (**B**) Superconducting gap (0.4 K; set point, 25 mV, 0.1 nA) at a location away from Se vacancies.

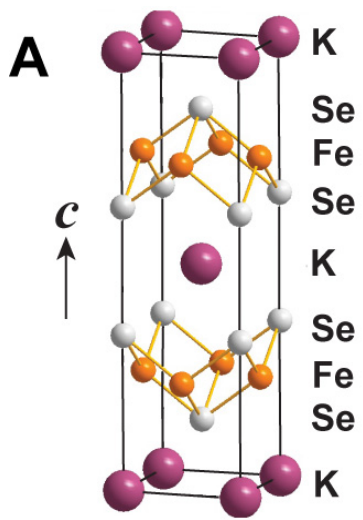 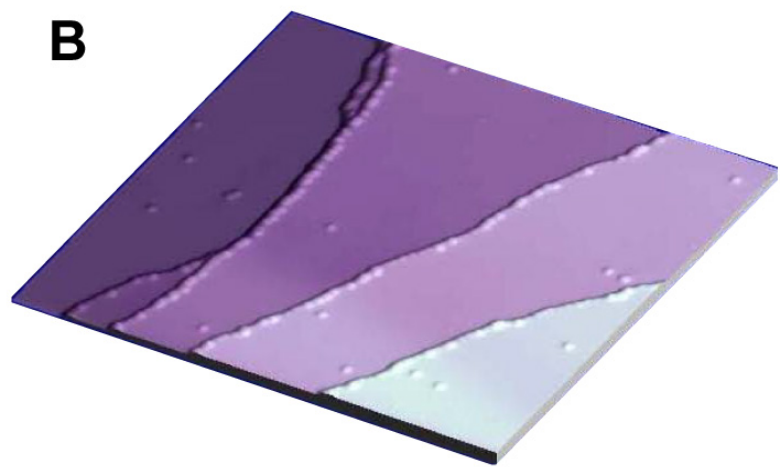 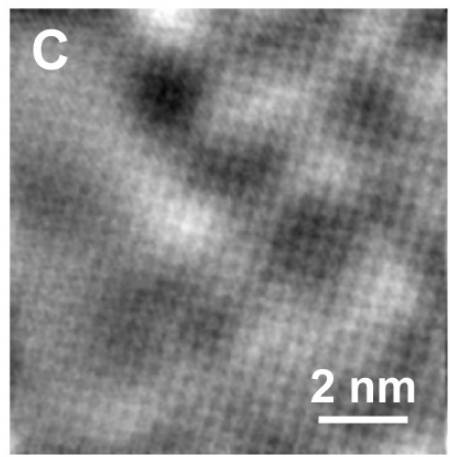 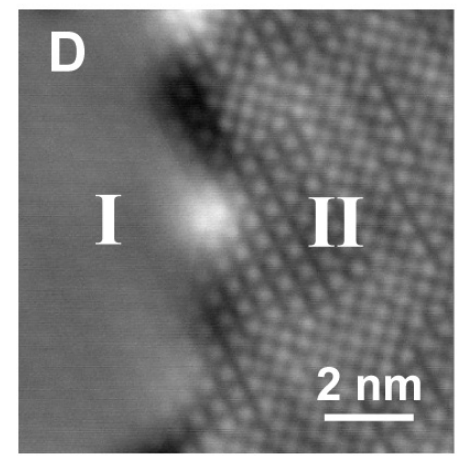
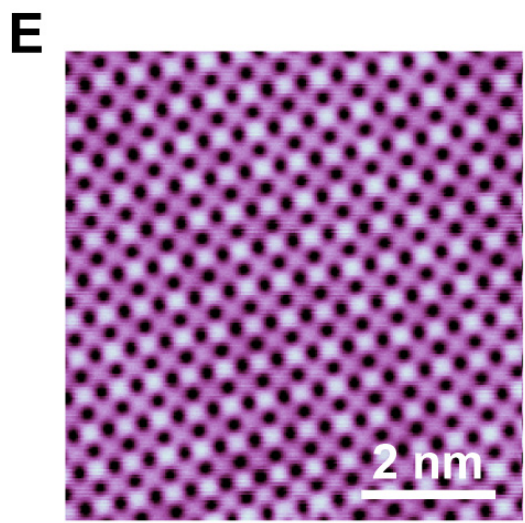 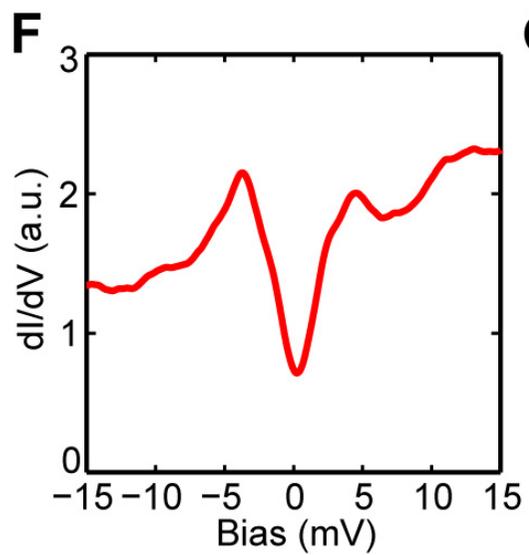 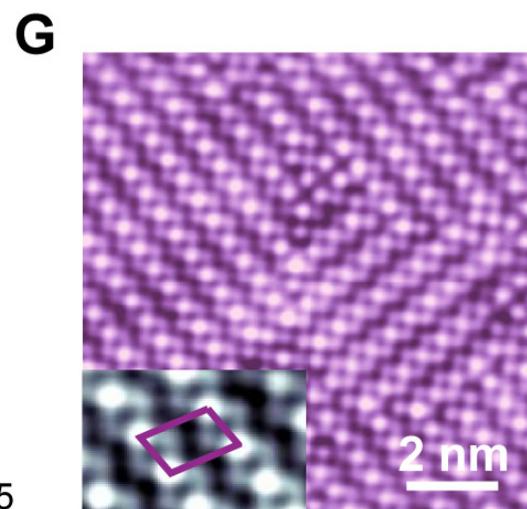 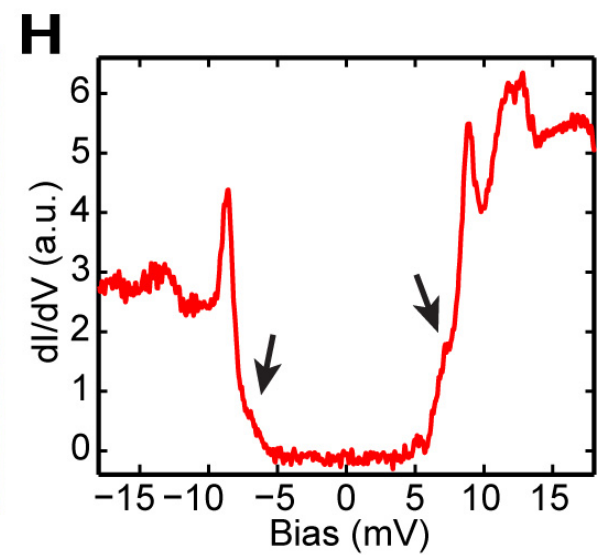

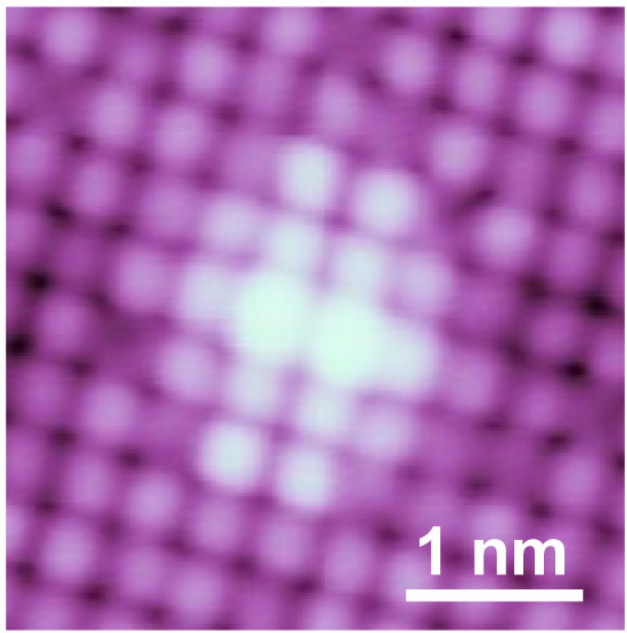
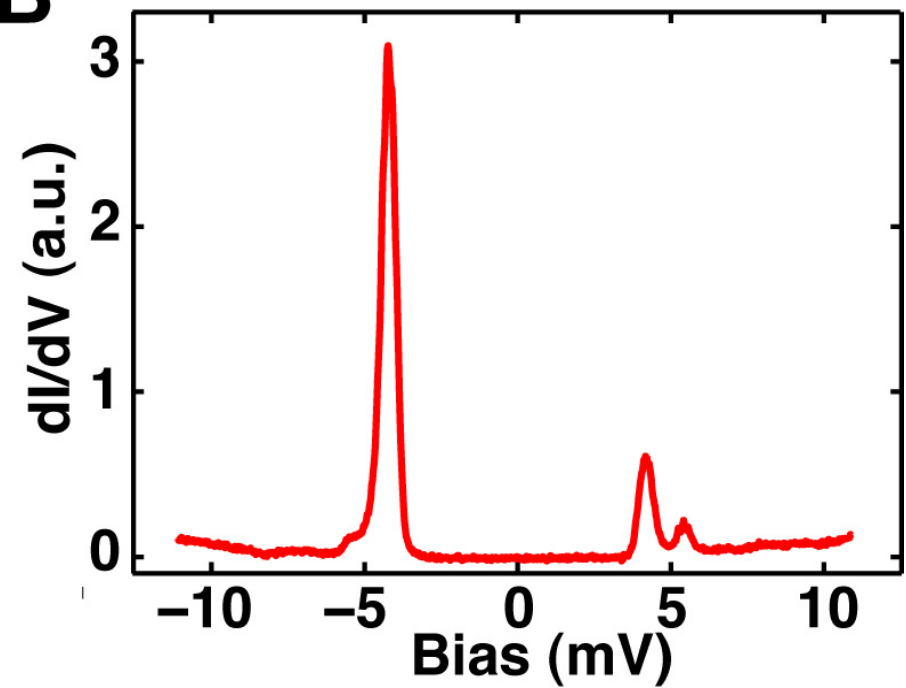
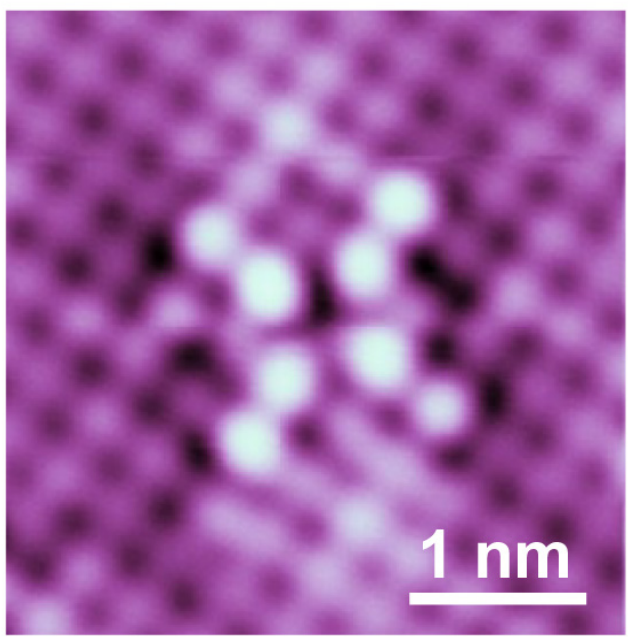
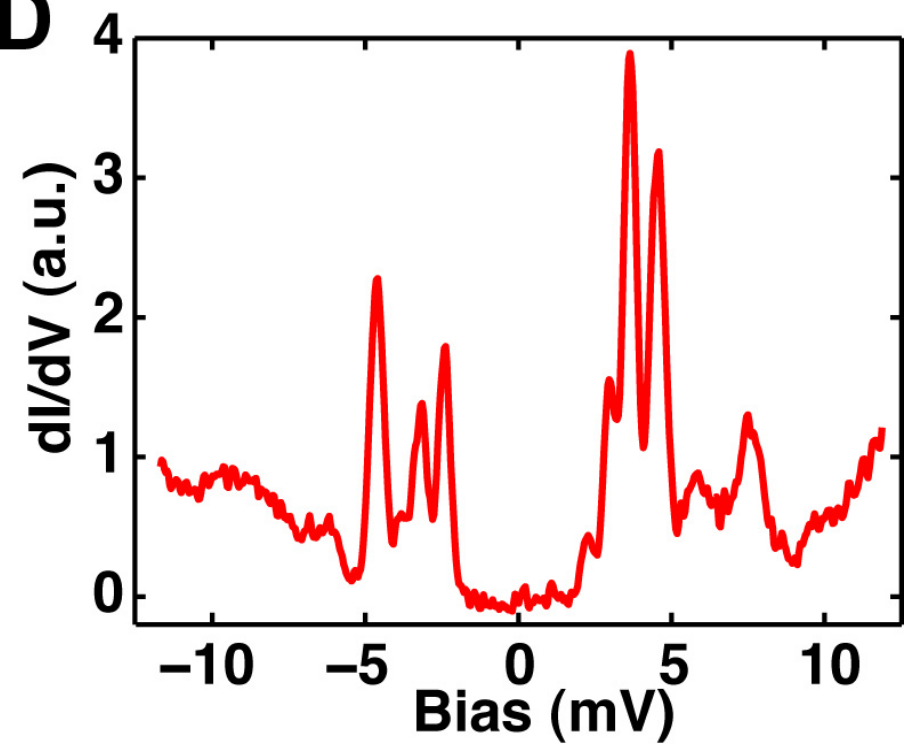

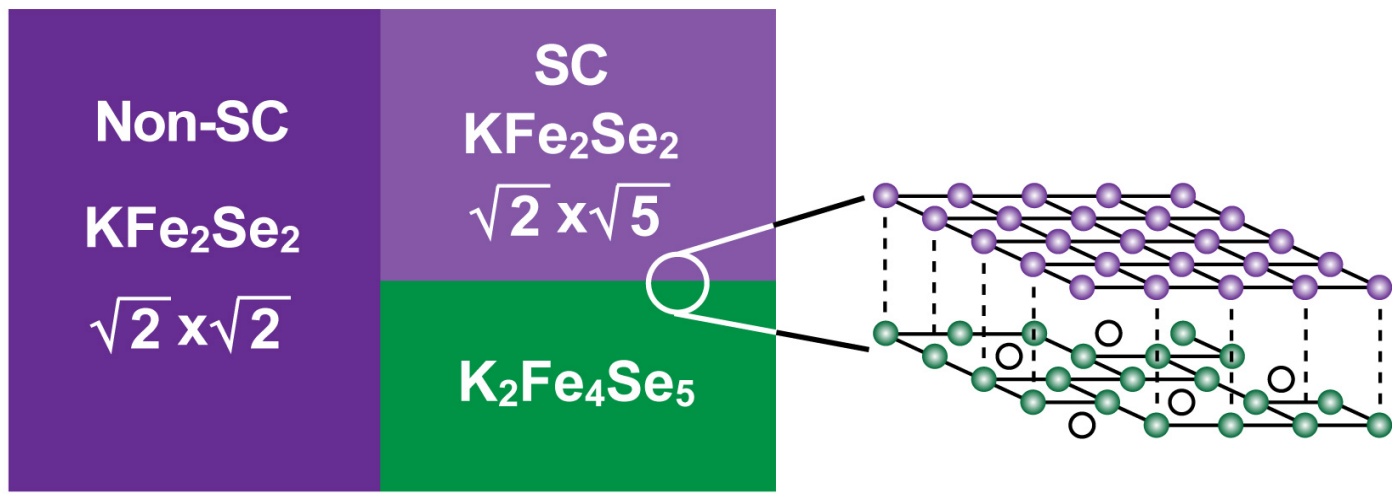

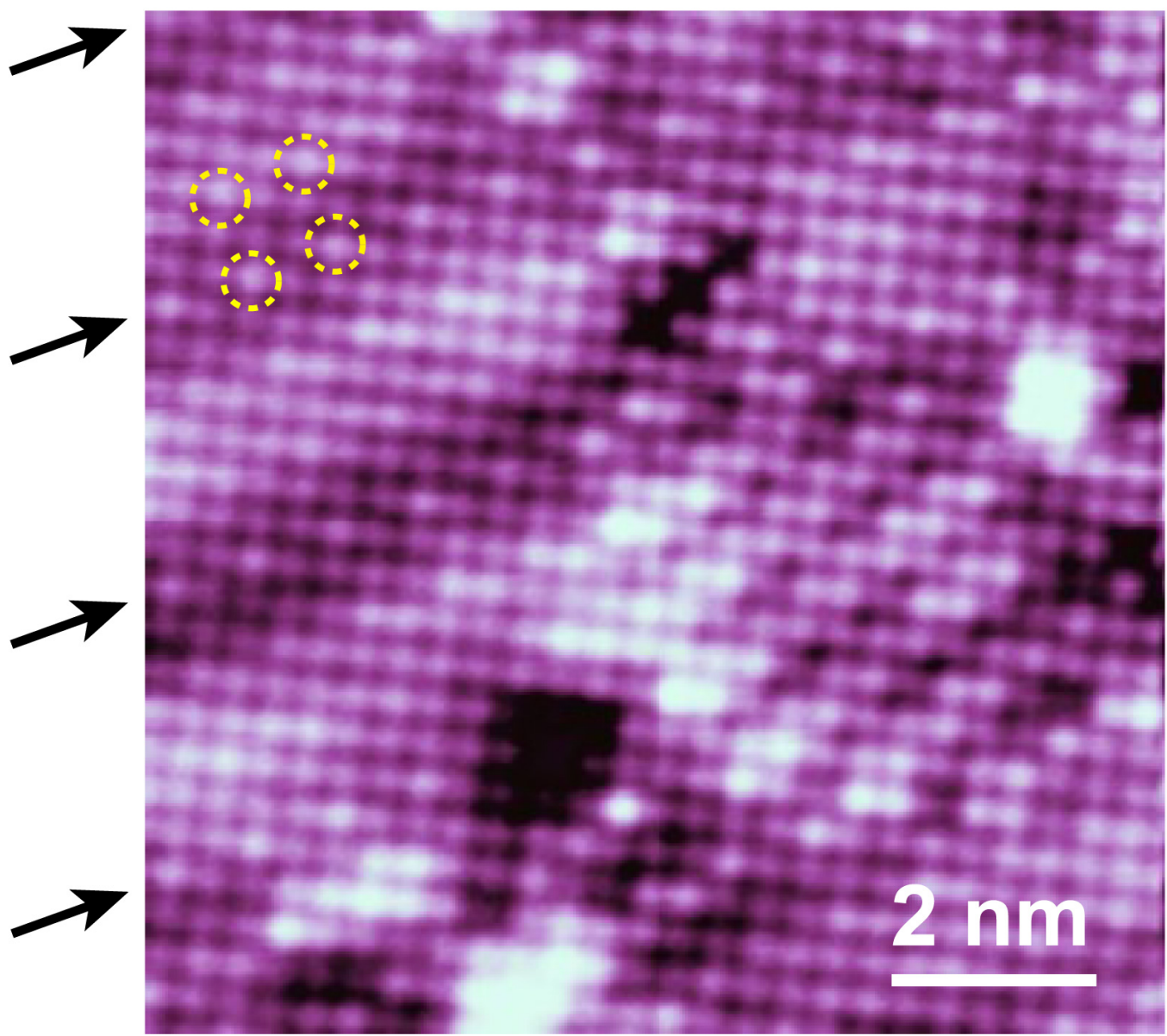

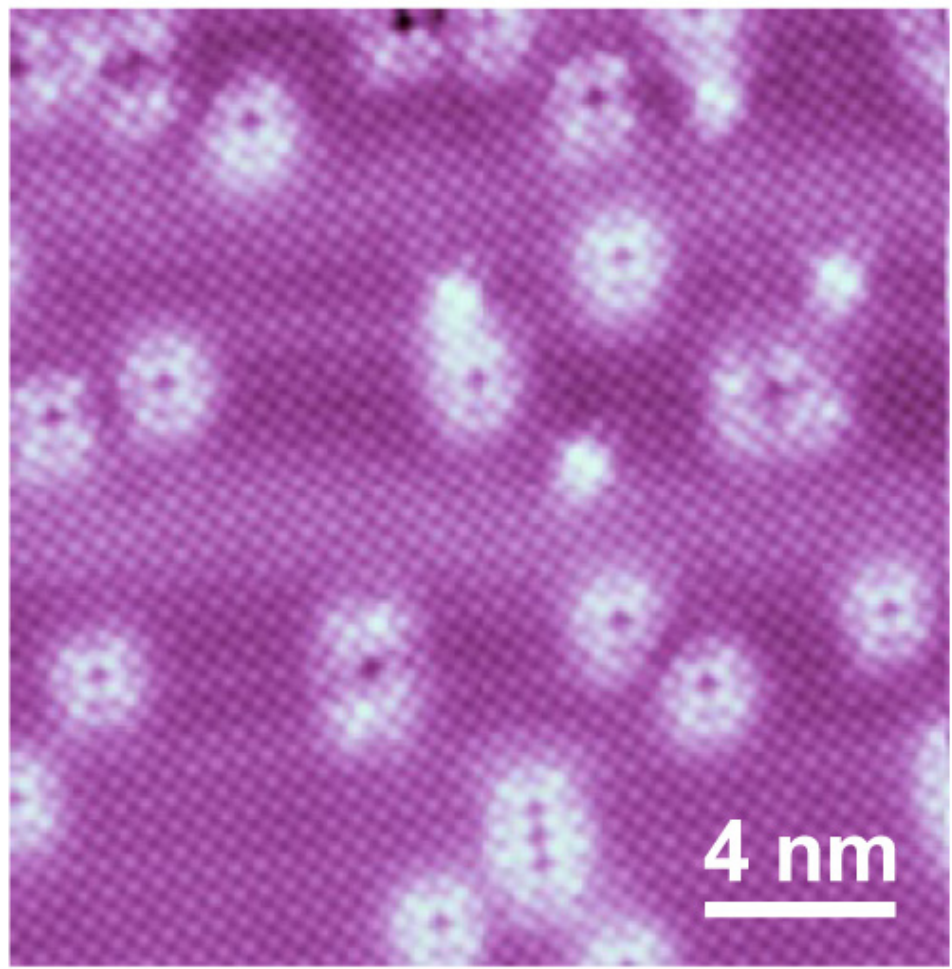 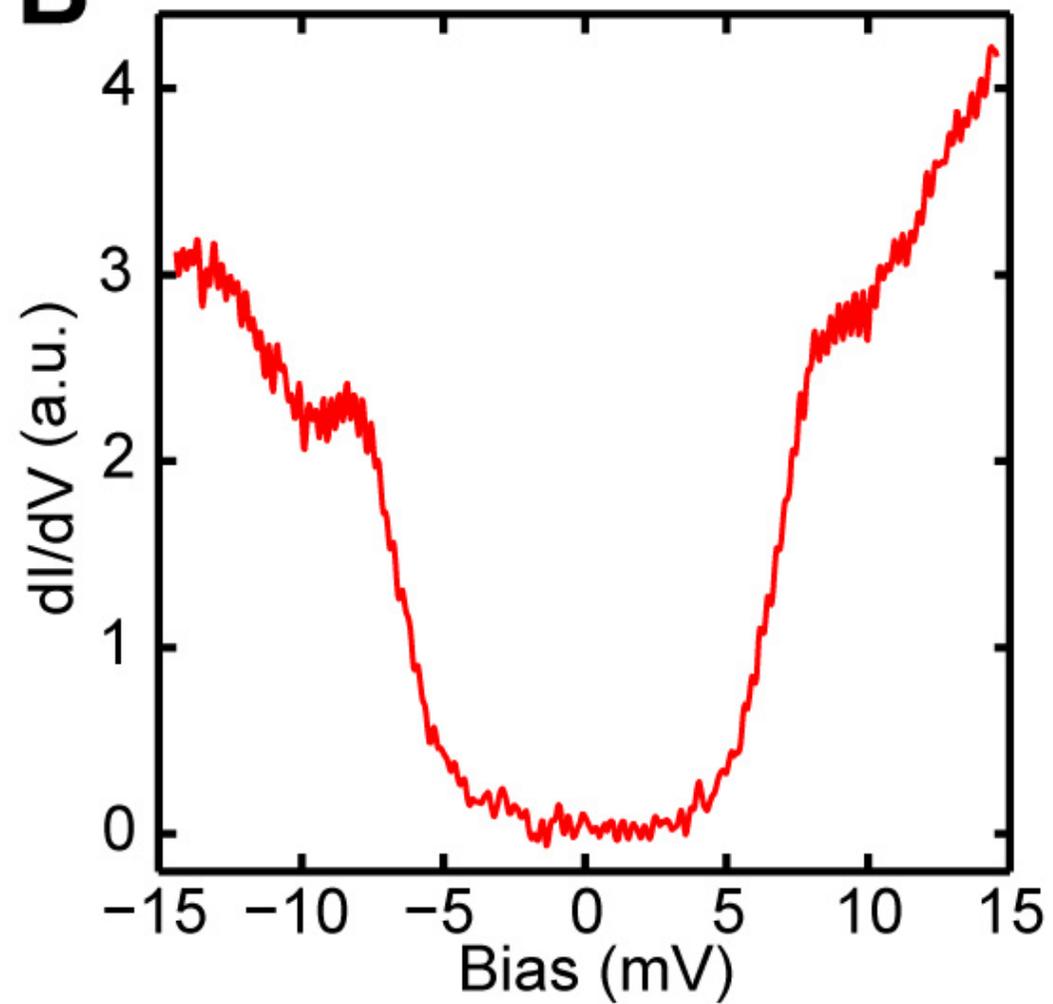

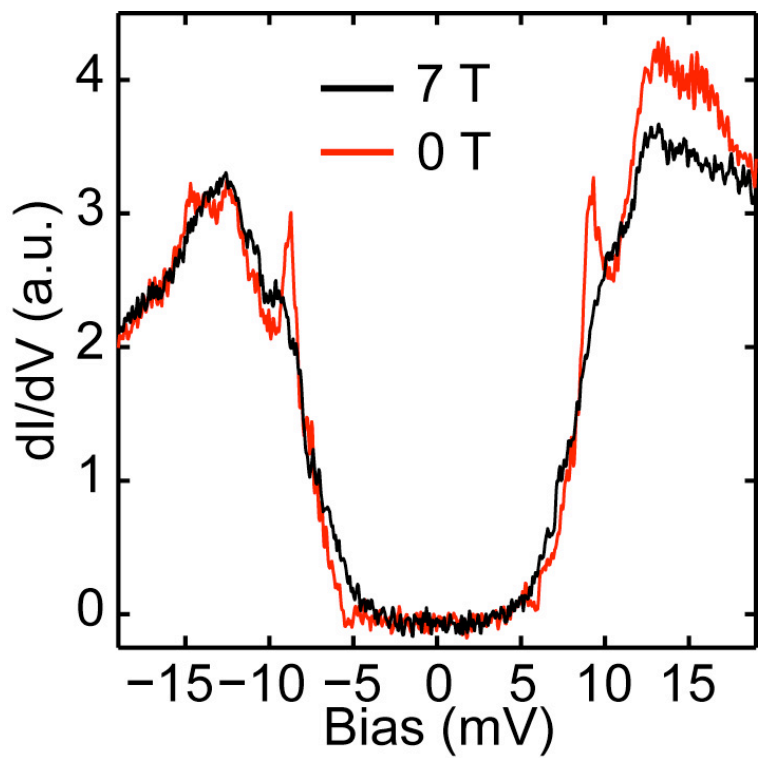

**Fig. S1.** Suppression of superconducting coherence peaks by magnetic field. Setpoint: 20 mV, 0.1 nA.

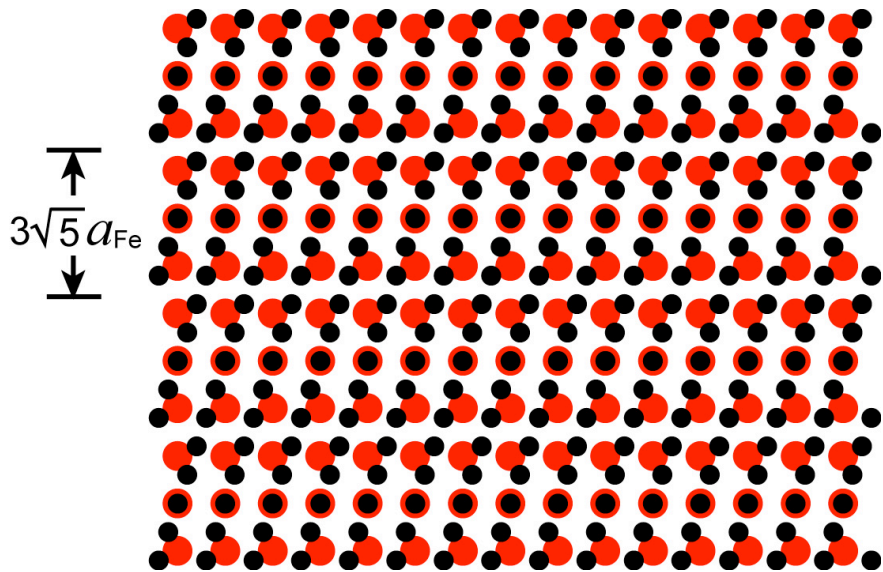

**Fig. S2.** Simulation of Morié pattern by superimposing the √2×√5 and √5×√5 lattices together. The period of the pattern is three times of that for the √5×√5 lattice.

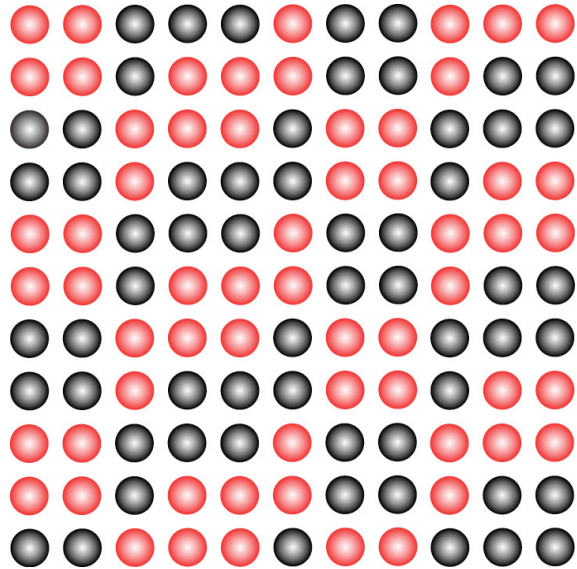

**Fig. S3.** A possible spin structure for the √2×√5 phase. The black and red colors represent two opposite spin directions. Such pattern can lead to √2×√5 superstructure in the topmost Se-layer.

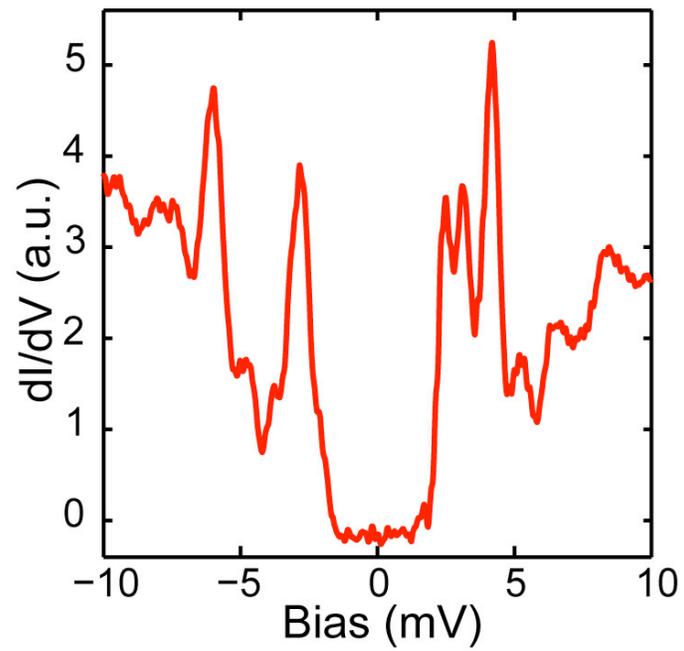

**Fig. S4.** Bound states induced by a single Se vacancy in superconducting $KFe_2Se_{2-z}$. Setpoint: 25 mV, 0.1 nA. The spectrum is similar to that in Fig. 2D.